# Electronic liquid crystals and superconductivity from doping a moiré Mott insulator

Yichi Zhang[1,2*], Yiyu Xia[2,3*], Zhongdong Han[1,2*], Jekwan Lee[1,2], Brandon Bauer[4], Kenji Watanabe[5], Takashi Taniguchi[5], Wenjin Zhao[4], Kin Fai Mak[1-3,6†], Jie Shan[1-3,6†]

[1]Laboratory of Atomic and Solid State Physics, Cornell University, Ithaca, NY, USA
[2]Max Planck Institute for the Structure and Dynamics of Matter, Hamburg, Germany
[3]School of Applied and Engineering Physics, Cornell University, Ithaca, NY, USA
[4]Department of Physics and Astronomy, University of Delaware, Newark, DE, USA
[5]National Institute for Materials Science, Tsukuba, Japan
[6]Kavli Institute at Cornell for Nanoscale Science, Ithaca, NY, USA

[*]These authors contributed equally
[†]Email: kin-fai.mak@mpsd.mpg.de; jie.shan@mpsd.mpg.de

**Electronic liquid crystals, such as stripe phases and nematics, are self-organized states of matter with spontaneously broken rotational symmetry[1-6]. They are expected to emerge in a doped Mott insulator as a compromise between short- and long-range Coulomb interactions[7,8], and are believed to be intimately connected to the emergence of high-temperature superconductivity either as a competing order or an intertwined order[9,10]. Moiré semiconductors provide a controllable platform to explore the relationship between stripe phases and superconductivity near a Mott insulator[11-21]. Here we report the generic appearance of stripe phases in a doped moiré Mott insulator and the observation of a stripe-to-superconductor quantum phase transition in twisted bilayer $WSe_2$. The stripe phases were revealed by electrical transport studies as a thermodynamic phase transition from a high-temperature isotropic phase to a low-temperature anisotropic phase accompanied by a Fermi surface reconstruction. Near the stripe-to-superconductor transition, we also observed an additional thermodynamic phase transition from metallic stripes to pinned insulating stripes. The stripe-to-superconductor transition further exhibits an intriguing self-similarity that highlights the intimate relationship between the two orders. Our results pave the path for studying intertwined orders[9] in moiré materials.**

## Main

Strongly correlated electronic systems are highly susceptible to charge localization and the formation of intermediate-scale structures[1-5,9,10,22-24]. In a doped antiferromagnetic Mott insulator, for example, the doped carriers and the antiferromagnetic domains like to phase separate, but the long-range Coulomb repulsion frustrates large-scale phase separation[1-4]; the compromise between these competing interactions is to form an intermediate-scale stripe phase with the doped carriers delocalized along the stripes, which are separated by intervening regions of nearly undoped antiferromagnetic domains[5,7,8] (Fig. 1a,b). The formation of such electronic stripes is well documented in high-temperature superconductors, such as the cuprates[9,24,25]; they appear near the superconducting domes, and are believed to have an intricate relationship with the superconducting state either as a competitor (in the case of static stripes)[26,27] or a promoter (in the case of fluctuating stripes)[28,29]. Understanding the connection between electronic stripes and superconductivity could be the key to unlock the mechanism for high-temperature superconductivity[9].

Moiré materials provide a platform to study this connection in a controllable manner[11-13,15,16,21,30-34]. In particular, moiré semiconductors have realized a highly tunable setup for exploring the physics of the extended Hubbard model[35]. Mott insulators[36,37], Wigner-Mott insulators[37,38] and kinetic magnetism[39,40] had been demonstrated in the strong coupling limit. In the intermediate coupling regime, superconductivity right next to a bandwidth-tuned Mott transition was observed in twisted bilayer $WSe_2$ ($tWSe_2$)[17-20]. When the Mott insulator barely survives at intermediate twist angles ($\theta \approx 4.6^0$), a complex phase diagram mimicking that of the cuprates had been reported[19]. A natural question arises: Are stripe phases ubiquitous near a moiré Mott insulator? If yes, are the stripe phases and superconductivity intimately connected? We attempt to address these questions in this study.

**Fermi surfaces of a doped Mott insulator**

We study dual-gated Hall bar devices of $tWSe_2$ with twist angles around $\theta \approx 4.0^0$. The top and bottom gates allow independent control of the hole doping density (measured by the lattice filling factor, $\nu$) in the material and the electric field ($E$) perpendicular to the sample plane. We also patterned contact and split gates, respectively, to lower the contact resistance and to remove unwanted parallel conduction channels for reliable four-terminal measurements of the longitudinal resistivity ($\rho_\parallel$), the magnetic field-symmetric transverse resistivity ($\rho_\perp$), and the magnetic field-antisymmetric Hall coefficient ($\rho_H$) as a function of $\nu$, $E$, temperature ($T$) and perpendicular magnetic field ($B$). Although we focus on a specific twist angle $\theta \approx 3.9^0$ in the main text, the reported phenomena below are reproducible in other twist angles (Extended Data Fig. 7 and 8). See Ref. [17,19] for details on device fabrication, Methods for electrical transport measurements, and Extended Data Fig. 1 for device schematic and image.

The topmost moiré valence band of $tWSe_2$, when non-topological (valid for $\theta \gtrsim 4.0^0$, Ref. [41]), can be mapped to a single-band moiré Hubbard model on a triangular lattice[42-45]. The bandwidth can be tuned, coarsely, by $\theta$ and, continuously, by $E$ (Ref. [19,20,46,47]). The electric field also tunes the band structure[43]: It introduces an effective spin-orbit coupling and tunes the spin-orbit splitting of the moiré band; it also tunes the location of the van Hove singularity (vHS) continuously (Fig. 1c and d, see Ref. [17,19,48,49] for details on band structure calculation). The combined actions of $\theta$ and $E$ thus provide an effective control of the coupling strength and the band structure in the moiré Hubbard model.

Figure 1e shows $\rho_\parallel$ of a $\theta \approx 3.9^0$ device as a function of $\nu$ and $E$ at $T \approx 20$mK and $B = 0$T. Consistent with earlier results[17-20], a higher $\rho_\parallel$ is generally observed in the layer-hybridized region, which is separated from the layer-polarized region by black dashed lines. The location of the vHS can be traced by a dip in $\rho_\parallel$, which is located near $\nu = 0.8$ at $E = 0$mV/nm and disperses to higher fillings with increasing $E$. Note that $\rho_{\parallel}$ at the vHS evolves from a peak to a dip at $T \lesssim 150\text{mK}$ (Extended Data Fig. 2). Due to the strong Hubbard interaction, the phase diagram in the layer-hybridized region at $\nu = 1$ is dominated by a Mott insulator exhibiting diverging $\rho_{\parallel}$ at low temperatures. A robust superconducting state with $\rho_{\parallel} = 0\Omega$ emerges in the vicinity of an electric field-tuned Mott transition near $E = \pm 25\text{mV/nm}$; the state spans over a region of $0.9 \lesssim \nu \lesssim 1.05$ and $7 \lesssim |E| \lesssim 25\text{mV/nm}$. In addition to the superconducting state, there are also regions (for both $\nu < 1$ and $\nu > 1$) surrounding the Mott insulator with higher $\rho_{\parallel}$ than the rest of the layer-hybridized region. These regions correspond to a doped Mott insulator with small Fermi surfaces.

To show the emergence of small Fermi surfaces surrounding the Mott insulator, we study the weak-field Hall response ($\rho_H$ measured at $B = 50\text{mT}$) as a function of $\nu$ and $E$ in Fig. 1f. We zoom into the doped Mott insulator region with $E > 0\text{mV/nm}$. Outside the doped Mott insulator region, $\rho_H$ is small, corresponding to a high doping density or a large Fermi surface. (Note the sign change in $\rho_H$ at the vHS near $\nu = 0.8$.) The Hall response or the Fermi surface here is band-like. In contrast, $\rho_H$ inside the doped Mott insulator region is substantially larger in general, which corresponds to a low doping density or a small Fermi surface. The emergence of a small Fermi surface is further demonstrated by a line cut at constant $E \approx 50\text{mV/nm}$ (Fig. 1h); it shows that the Hall density ($n_H$) is referenced to doping from $\nu = 1$ in this region; it also shows a jump in $n_H$ by the moiré density ($n_M$) at the boundary separating the doped Mott insulator and band-like regions.

We summarize the Fermi surfaces in the vicinity of the $\nu = 1$ Mott insulator in Fig. 1g. A small electron (hole) Fermi surface dominates the phase space for $\nu < 1$ ($\nu > 1$), although small regions hosting hole Fermi surfaces can also be identified at $\nu < 1$. The results demonstrate the survival of the Mott gap upon doping the Mott insulator away from $\nu = 1$, which results in small electron/hole Fermi surfaces; sufficiently strong doping eventually quenches the Mott gap and drives a transition to band-like Fermi surfaces. The critical doping density this transition happens is electric field dependent as the Mott gap size depends on $E$. The results are consistent with the published data at other twist angles[17,19,20].

**Generic emergence of stripe phases**

We now examine the magnetic field-symmetric transverse response, $\rho_{\perp}$, as a function of $\nu$ and $E$ (at $T \approx 20\text{mK}$ and $B = 0\text{T}$) for two different transverse electrode pairs in the Hall bar device (Fig. 2a,b). The response is sensitive to the emergence of transport anisotropy[33,34]. Whereas a negligible $\rho_{\perp}$ is observed outside the doped Mott insulator (i.e. band-like) region, a large transverse response is generally observed inside. Variations in $\rho_{\perp}$ (in both magnitude and sign) are also observed between $E > 0$ and $E < 0$ and among the different transverse electrode pairs. Moreover, a sign change in $\rho_{\perp}$ is observed under a rotation in the probe direction by 90 degrees (Extended Data Fig. 3).

To further investigate the emergent transport anisotropy in the doped Mott insulator region, we study the temperature dependence at the two line cuts marked in Fig. 2a. Figures 2d-f show, respectively, $\rho_{\parallel}$, the anisotropy ratio $\frac{\rho_{\perp}}{\rho_{\parallel}}$ and $\rho_H$ (measured at $B = 100\text{mT}$) as a function of $\nu$ and $T$ at constant $E \approx 54\text{mV/nm}$. The corresponding plots as a function of $E$ and $T$ at constant $\nu = 0.915$ are shown in Fig. 2g-i. We also show the temperature dependence of $\rho_{\parallel}$, $\frac{\rho_{\perp}}{\rho_{\parallel}}$ and $\rho_H$ at selected $\nu$ and $E$ in Fig. 3. The results are insensitive to thermal cycling. A peak/bump in $\rho_{\parallel}$ is observed at a transition temperature $T_S \approx (1-6)\text{K}$ (dependent on $\nu$ and $E$), below which both $\frac{\rho_{\perp}}{\rho_{\parallel}}$ and $\rho_H$ increase from a small value. (The small but nonzero $\frac{\rho_{\perp}}{\rho_{\parallel}}$ above $T_S$ is caused by an unintentional longitudinal-transverse mixing in the device due to the imperfect Hall bar geometry.) The dependence of $T_S$ on $\nu$ and $E$ is summarized in Fig. 2d-f and Fig. 2g-i, respectively. The Mott insulator appears below $T \approx 8\,\text{K}$; $T_S$ vanishes continuously from this Mott scale as doping increases to ≈18% from $\nu = 1$. Meanwhile, the electric field dependence of $T_S$ shows a dome shape.

The results above demonstrate strong experimental evidence of an emerging stripe phase at $T < T_S$ by doping a Mott insulator. First, the temperature dependent transport anisotropy $\frac{\rho_{\perp}}{\rho_{\parallel}}$ supports a thermodynamic phase transition from an isotropic to a stripe phase at $T_S$. The stripe direction is in general not perfectly aligned parallel or perpendicular to the major axis of the Hall bar device, thus giving rise to a finite transverse response $\frac{\rho_{\perp}}{\rho_{\parallel}}$ at $T < T_S$ ($\frac{\rho_{\perp}}{\rho_{\parallel}}$ is a proxy for the stripe order parameter, see Methods). The variations in both the magnitude and sign of $\frac{\rho_{\perp}}{\rho_{\parallel}}$ between $E > 0$ and $E < 0$ and among the different transverse electrode pairs (Fig. 2a, b) show that the overall stripe direction can, in general, vary spatially in the micron-scale device channel and with gating. (Note that micron-scale stripe domains had been reported in angle-aligned $WSe_2/WS_2$ moiré bilayers[12].) The local stripe direction is likely oriented by the local unintentional strain in the device channel[50], as supported by the observed insensitivity to thermal cycling.

Second, the observed decrease in Hall density below $T_S$ shows that the transport anisotropy is accompanied by a Fermi surface reconstruction. The result is inconsistent with a nematic phase without translational symmetry breaking, in which a change in Hall density is not expected; rather, the reconstructed Fermi surface supports the appearance of a new super-period for zone folding and thus the emergence of a stripe phase with translational symmetry breaking (e.g. smectic phase). The Fermi surface reconstruction is also revealed by the peak/bump in $\rho_{\parallel}$ at $T_S$ as parts of the large Fermi surface are gapped out. Similar phenomenology had been reported in the cuprate high-temperature superconductors[51] and in other density wave materials[52].

Third, the stripe phase emerges from doping a moiré Mott insulator, as suggested by Ref. [15]. The intimate connection to the Mott insulator is supported by 1) the presence of a substantial transport anisotropy only in the doped Mott insulator region with a small Fermi surface, and 2) the systematic evolution of $T_S$ from the Mott insulator temperature scale ($\approx$ 8K) as doping increases from $\nu = 1$. See Methods and Extended Data Fig. 3 and 4 for additional evidence for the emergence of stripe phases.

**Pinned insulating stripes**

In addition to the phase transition from an isotropic metal to a stripe metal at $T_S$, we also identify a metal-to-insulator transition over the range $25 \lesssim E \lesssim 40$ mV/nm at a lower transition temperature $T_S' \approx (0.1 - 0.3)$K, below which $\rho_\parallel$ increases substantially (Fig. 2d and Fig. 3a); a peak in $\frac{\rho_\perp}{\rho_\parallel}$ is also observed at $T_S'$ (Fig. 4a). (Note that $\rho_H$ becomes noisy below $T_S'$ due to the high sample resistance). The result suggests a two-step phase transition over the range $25 \lesssim E \lesssim$ 40mV/nm: first from an isotropic metal to a stripe metal at $T_S$, followed by another transition from a metallic to an insulating stripe phase at $T_S'$.

To confirm the emergence of an insulating stripe phase below $T_S'$, we perform bias dependence studies to unveil its depinning. Figure 4b shows the differential longitudinal resistance ($\frac{dV_\parallel}{dI}$) as a function of the DC bias current ($I$) and $T$ at $\nu = 0.915$ and $E = 36$mV/nm ($V_\parallel$ denotes the longitudinal voltage drop). The zero-bias differential resistance $\left.\frac{dV_\parallel}{dI}\right|_{I=0}$ shows a metal-to-insulator transition at $T_S' \approx 200$mK as discussed above. Below $T_S'$, a highly non-ohmic transport behavior emerges: A small bias current dramatically suppresses the zero-bias resistance peak and gives a negative differential resistance around $I = 3$nA and, finally, ohmic transport for $I \gtrsim 6$nA. (See Extended Data Fig. 5 for similar behaviors for the differential transverse resistance.) The non-ohmic transport is further illustrated by the temperature dependent $I - V_\parallel$ characteristics in Fig. 4c. For $T < T_S'$, we observed a highly nonlinear $I - V_\parallel$ dependence exhibiting a threshold onset near $V_\parallel \approx \pm 10\mu$V and a S-shape dependence (and thus a hysteresis) near the threshold; the S-shape dependence corresponds to the negative differential resistance in Fig. 4b.

The observed non-ohmic transport at $T < T_S'$ is fully consistent with the behavior of sliding density wave transport induced by depinning of the insulating stripes near the threshold bias[53,54]. The dependence can be understood by a two-fluid model consisting of 1) normal single-particle transport due to thermal activations and 2) collective transport of a sliding (stripy) density wave condensate for bias above the threshold[53]. The hysteresis or the negative differential resistance reflects the different bias thresholds for density wave depinning with increasing bias and for re-trapping of the density wave by the pinning potential with decreasing bias[53]. The results provide a basis for future studies of the dynamics of collective (stripy) density wave transport in the time/frequency domain[53].

We summarize the phase diagram for stripe phases in Fig. 2c. Comparison with Fig. 1g shows that the stripe phases occupy a substantial portion (but not all) of the doped Mott insulator region. In addition to stripe metals, a pinned insulating stripe phase is also identified over a region of $0.92 \lesssim \nu \lesssim 0.98$ and $25 \lesssim |E| \lesssim 40$mV/nm. (See Extended Data Fig. 6 for additional sliding density wave transport data in this region of the phase space.)

**Insulating stripe-to-superconductor transition**

Finally, we investigate the electric field-induced insulating stripe-to-superconductor transition. Figure 5a shows a zoom-in view of $\rho_{\parallel}$ as a function of $E$ and $T$ at constant $\nu \approx 0.94$ near the critical electric field $E_C \approx 24.2$mV/nm; the electric field dependence of the superconducting $T_C$ and $T_S'$ are also shown. Line cuts at selected electric fields near $E_C$ are shown in Fig. 5c. The continuously vanishing $T_C$ and $T_S'$ towards the critical point at $E_C$ suggests that the quantum phase transition is continuous in nature. This is further evidenced by a scaling collapse of the resistivity curves[17] (see Methods for details). By normalizing the resistivity curves $\rho_{\parallel}(T)$ at electric fields away from $E_C$ by that at $E_C$ (denoted as $\rho_C(T)$, dashed curve in Fig. 5c) and by scaling the temperature $T$ by an electric field dependent temperature scale $T_0$, the scaled resistance curves can be collapsed into two groups: one on the superconductor side and another on the insulator side (Fig. 5d). The result implies that the resistivity curves near $E_C$ have a universal form $\rho(E,T) = \rho_C(T)\tilde{f}[(E-E_C)/T^{1/z\nu}]$ ($\tilde{f}$ is a universal function). The extracted exponent is $z\nu \approx 1$ on both sides of the transition (inset of Fig. 5d).

We construct the phase diagram near $E_C$ in Fig. 5b. For $E < E_C$ and away from the critical point, a Fermi liquid behavior, $\rho_{\parallel} \propto T^2$, emerges at temperatures below the coherence temperature ($T^*$). We extracted $T^*$ as a function of the electric field in Fig. 5a,b (Methods); it vanishes continuously towards the critical point, i.e. the $\frac{T_C}{T^*}$ ratio increases towards the critical point. In addition, a non-Fermi liquid behavior, as signified by a $T$-linear temperature dependence for $\rho_{\parallel}$, is observed in the immediate vicinity of the critical point. For $E > E_C$, the phase diagram is dominated by the stripe metal phase at $T > T_S'$.

The phase diagram also shows 1) a comparable scale and 2) a similar electric-field dependence with reference to $E_C$ for both $T_C$ and $T_S'$. The result suggests an intimate connection between the superconductor and the insulating stripe phase, and that the two orders may be intertwined[9]. A possible microscopic mechanism for the observed insulating stripe-to-superconductor transition is that lowering the electric field increases the transverse fluctuations of the stripes and thus suppresses the tendency towards the stripy density wave order and favors the superconducting instability[5,7]. Demonstration of a pair density wave[9] in the superconducting region right next to the insulating stripe region could provide direct evidence for intertwined ordering in future experiments.

In conclusion, we demonstrate a generic emergence of stripe phases by doping a moiré Mott insulator and the observation of a stripe-to-superconductor quantum phase transition in $tWSe_2$. Right next to this transition (Fig. 5b), a two-step thermodynamic phase transition first from an isotropic metal to a stripe metal, followed by another from a metallic to an insulating stripe, is observed. The major results of this study are reproducible in other devices with different twist angles (Extended Data Fig. 7 and 8), suggesting that stripe phases are universal in a doped moiré Mott insulator. Future studies are required to further illustrate the connection between electronic stripes and superconductivity.

## Methods

### Electrical transport measurements

Electrical transport measurements were performed in a dilution refrigerator (Bluefors LD250). Low-temperature RC (resistor-capacitor) and RF (radiofrequency) filters (QDevil) were installed on the mixing chamber to filter out unwanted electrical signals with frequencies above ≈65kHz. A room-temperature resistor (5MΩ), much larger than the sample and contact resistances at most gate voltages, was connected in series with the device for current biasing. Low noise voltage preamplifiers (Stanford Research Systems SR560) with large input impedance (100MΩ) and current preamplifiers (Ithaco Model 1211) were used to measure the sample four-terminal resistance. We adapted the low-frequency (7.137Hz) lock-in technique to measure the sample resistance using a small excitation current ($\lesssim$ 1nA) to minimize high-bias effects. We obtained the longitudinal resistivity, $\rho_\parallel = \frac{R_\parallel W}{L}$, using the measured longitudinal resistance $R_\parallel$ and the device dimension ratio $\frac{W}{L} \approx 0.77$. The Hall coefficient, $\rho_H = \frac{\rho_{xy}}{B}$, was obtained from the antisymmetrized transverse resistance, $\rho_{xy} = [R_{xy}(B) - R_{xy}(-B)]/2$, measured under a small magnetic field $B =$ 50 or 100mT. The field-symmetric transverse resistivity, $\rho_\perp = [R_{xy}(B) + R_{xy}(-B)]/2$, is approximately equal to the zero-field transverse resistivity, $\rho_\perp \approx R_{xy}(B = 0)$.

### Anisotropic transport probed by transverse resistivity

We probed the stripe phase by measuring the anisotropic resistivity tensor in the Hall bar device, following the methodology established by previous studies[33,55]. In a typical device, the current direction is misaligned with the principal axes of the system (in our case, the single ordering vector $\boldsymbol{q}$ of the stripe phase) by an angle $\phi$. The resistivity tensor acquires an off-diagonal term:

$$\rho = \begin{pmatrix} \frac{1}{2}(\rho_1 + \rho_2) + \frac{1}{2}(\rho_1 - \rho_2)\cos 2\phi & \frac{1}{2}(\rho_1 - \rho_2)\sin 2\phi \\ \frac{1}{2}(\rho_1 - \rho_2)\sin 2\phi & \frac{1}{2}(\rho_1 + \rho_2) - \frac{1}{2}(\rho_1 - \rho_2)\cos 2\phi \end{pmatrix}, \quad (1)$$

where $\rho_1, \rho_2$ are the resistivity tensor elements measured at the principal axes direction. In contrast to the Hall resistivity, the off-diagonal term here ($= \rho_\perp$) is even under magnetic field and in general nonzero at $B = 0$T. According to equation (1), a sign change in $\rho_\perp$ is expected under a rotation in the probe direction by 90 degrees. This is verified in Extended Data Fig. 3, which compares $\rho_\perp$ probed at two directions rotated by 90 degrees.

**Extraction of temperature scales**

We defined four different temperature scales in this study: the superconducting transition temperature $T_C$, the stripe metal transition temperature $T_S$, the stripe insulator transition temperature $T_S'$, and the Fermi liquid coherence temperature scale $T^*$. The superconducting $T_C$ was defined as the temperature at 50% deviation from the normal state resistivity. As shown in the main text (e.g. Fig. 3), $\rho_\parallel$, $\frac{\rho_\perp}{\rho_\parallel}$ and $\rho_H$ all show characteristic changes at $T_S$. We can therefore determine $T_S$ as the temperature at the $\rho_\parallel$ peak/bump or as the onset temperature for $\frac{\rho_\perp}{\rho_\parallel}$ and $\rho_H$. The results are all consistent. In the main text, we show results obtained from $\rho_H$. For $T_S'$, we used the peak in $\frac{\rho_\perp}{\rho_\parallel}$ to determine its value; the value well correlates with the onset temperature for the insulating behavior in $\rho_\parallel$ (Fig. 4a).

To determine $T^*$ in Fig. 5a, we first note that the temperature dependent resistivity for $E \lesssim E_C$ can be described by $\rho_\parallel(T) = \rho_0 + AT^{\alpha(T)}$ when a magnetic field $B = 100$mT is applied to suppress superconductivity. Here $\rho_0$ is the residual resistivity (independent of $T$), $A$ is a temperature-independent coefficient, and $\alpha(T)$ is a temperature-dependent exponent ($\alpha = 2$ for Fermi liquid). The exponent $\alpha(T)$ is equal to the local slope of $\ln(\rho_\parallel - \rho_0)$ versus $\ln(T)$; $\rho_0$ was extracted as the y-intercept from a polynomial fit to the $\rho_\parallel$ data from 10mK to 2K. In general, $\alpha(T)$ decreases from $\alpha = 2$ to $\alpha = 1$ with increasing temperature (Extended Data Fig. 9). We define the crossover temperature scale $T^*$ as the temperature where $\alpha(T) < 1.5$.

**Further confirmation of anisotropic stripe phase**

Whereas the main text provides strong experimental evidence for the emergence of stripe phases through the systematic temperature and gate dependences of $\rho_\parallel$, $\frac{\rho_\perp}{\rho_\parallel}$ and $\rho_H$ and through the near perfect correlations between these measurable quantities, we performed additional experiments to rule out other mundane sources for transverse resistivity, such as electronic inhomogeneities as proposed in Ref. [56]. First, we rotate the probe direction in our four-terminal measurements by 90 degrees. According to equation (1), a sign change in $\rho_\perp$ is expected: $\rho_\perp\left(\phi + \frac{\pi}{2}\right) = \frac{1}{2}(\rho_1 - \rho_2)\sin[2\left(\phi + \frac{\pi}{2}\right)] = -\rho_\perp(\phi)$. This is exactly observed in Extended Data Fig. 3; moreover, the magnitude of $\rho_\perp$ remains nearly unchanged under the rotation. The result verifies equation (1) expected for a stripe phase, but is inconsistent with the origin from electronic inhomogeneities, in which no correlation between the sign and magnitude of $\rho_\perp$ is expected under a 90-degree rotation.

Second, in the trivial case of electronic inhomogeneities[56], a large $\rho_\perp$ is expected where $\rho_\parallel$ shows a strong dependence on the tuning parameters ($\nu$, $E$, etc.). To rule this out, we measured the two-terminal resistances, $R_{2t}$, along the longitudinal and transverse directions at three electric field line cuts in Extended Data Fig. 4. We chose line cut 1 through the stripe phases at constant $\nu = 0.937$ and line cuts 2 and 3 at constant $\nu = 0.175$ and $\nu = 0.2$, respectively, where there is no stripe phase but $\rho_\parallel$ shows a strong dependence on $E$. A pronounced two-terminal transport anisotropy is observed only for line cut 1 but not for 2 and 3. The result further verifies the emergence of stripe phases in a doped Mott insulator rather than electronic inhomogeneities. The electronic inhomogeneities scenario is also inconsistent with the systematic temperature and gate dependences of $\rho_\parallel$, $\frac{\rho_\perp}{\rho_\parallel}$ and $\rho_H$ and the near perfect correlations between these quantities as observed in our experiment.

**Analysis of critical scaling**

The scaling collapse is a characteristic behavior of continuous phase transitions. The critical field $E_C \approx 24.2\text{mV/nm}$ and the corresponding resistivity $\rho_C(T)$ were identified from the temperature dependent resistivity in Fig. 5a,c. To conduct scaling analysis, we first chose a group of resistivity curves at different electric fields in the immediate vicinity of $E_c$, and normalized these curves by $\rho_C(T)$. We assigned each curve a characteristic temperature scale $T_0$, and set $T_0 = T_C$ (or $T_0 = T_S'$) at the field furthest from $E_c$ (i.e. the endpoints). (Note that the magnitude of $T_0$ is arbitrary in the scaling analysis.) The value of $T_0$ for each normalized resistivity curve was determined using the least squares method to best collapse it onto the endpoint curve ($E = 22.2\text{mV/nm}$ on the superconductor side and $E = 26.2\text{mV/nm}$ on the stripe insulator side, see Fig. 5d). Extra caution was taken for the curves on the stripe insulator side, where a peak/hump at $T_S$ might cause errors in the determination of $T_0$. However, we found the least squares method is insensitive to the peak/hump feature at $T_S$ as the sum of the squared residuals is mainly contributed by the low temperature data with high resistivities. The scaling analysis yields a power law for $T_0$: $T_0 \propto |E - E_C|^{z\nu}$, where $z$ is the dynamical critical exponent and $\nu$ is the correlation length exponent (inset of Fig. 5d).

**References**

1 Zaanen, J. & Gunnarsson, O. Charged magnetic domain lines and the magnetism of high-Tc oxides. *Physical Review B* **40**, 7391-7394 (1989).

2 Machida, K. Magnetism in La2CuO4 based compounds. *Physica C: Superconductivity* **158**, 192-196 (1989).

3 Kato, M., Machida, K., Nakanishi, H. & Fujita, M. Soliton Lattice Modulation of Incommensurate Spin Density Wave in Two Dimensional Hubbard Model -A Mean Field Study-. *Journal of the Physical Society of Japan* **59**, 1047-1058 (1990).

4 Emery, V. J. & Kivelson, S. A. Frustrated electronic phase separation and high-temperature superconductors. *Physica C: Superconductivity* **209**, 597-621 (1993).

5 Emery, V. J., Kivelson, S. A. & Tranquada, J. M. Stripe phases in high-temperature superconductors. *Proceedings of the National Academy of Sciences* **96**, 8814-8817 (1999).
6 Fradkin, E., Kivelson, S. A., Lawler, M. J., Eisenstein, J. P. & Mackenzie, A. P. Nematic Fermi Fluids in Condensed Matter Physics. *Annual Review of Condensed Matter Physics* **1**, 153-178 (2010).
7 Kivelson, S. A., Fradkin, E. & Emery, V. J. Electronic liquid-crystal phases of a doped Mott insulator. *Nature* **393**, 550-553 (1998).
8 Löw, U., Emery, V. J., Fabricius, K. & Kivelson, S. A. Study of an Ising model with competing long- and short-range interactions. *Physical Review Letters* **72**, 1918-1921 (1994).
9 Fradkin, E., Kivelson, S. A. & Tranquada, J. M. Colloquium: Theory of intertwined orders in high temperature superconductors. *Reviews of Modern Physics* **87**, 457-482 (2015).
10 Fernandes, R. M., Orth, P. P. & Schmalian, J. Intertwined Vestigial Order in Quantum Materials: Nematicity and Beyond. *Annual Review of Condensed Matter Physics* **10**, 133-154 (2019).
11 Andrei, E. Y. *et al.* The marvels of moiré materials. *Nature Reviews Materials* **6**, 201-206 (2021).
12 Jin, C. *et al.* Stripe phases in WSe2/WS2 moiré superlattices. *Nature Materials* **20**, 940-944 (2021).
13 Kennes, D. M. *et al.* Moiré heterostructures as a condensed-matter quantum simulator. *Nature Physics* **17**, 155-163 (2021).
14 Mak, K. F. & Shan, J. Semiconductor moiré materials. *Nature Nanotechnology* **17**, 686-695 (2022).
15 Wietek, A. *et al.* Tunable stripe order and weak superconductivity in the Moiré Hubbard model. *Physical Review Research* **4**, 043048 (2022).
16 Nuckolls, K. P. & Yazdani, A. A microscopic perspective on moiré materials. *Nature Reviews Materials* **9**, 460-480 (2024).
17 Xia, Y. *et al.* Superconductivity in twisted bilayer WSe2. *Nature* **637**, 833-838 (2025).
18 Guo, Y. *et al.* Superconductivity in 5.0° twisted bilayer WSe2. *Nature* **637**, 839-845 (2025).
19 Xia, Y. *et al.* Bandwidth-tuned Mott transition and superconductivity in moiré WSe2. *Nature* **650**, 585-591 (2026).
20 Guo, Y. *et al.* Angle evolution of the superconducting phase diagram in twisted bilayer WSe2. *Nature* **652**, 622-627 (2026).
21 Erekle Jmukhadze, S. O., Allan H. MacDonald, Wei-Cheng Lee. Stripe antiferromagnetism and chiral superconductivity in tWSe2. *arXiv:2601.20836* (2026).
22 Dagotto, E. Complexity in Strongly Correlated Electronic Systems. *Science* **309**, 257-262 (2005).
23 Paschen, S. & Si, Q. Quantum phases driven by strong correlations. *Nature Reviews Physics* **3**, 9-26 (2020).
24 Lee, P. A., Nagaosa, N. & Wen, X.-G. Doping a Mott insulator: Physics of high-temperature superconductivity. *Reviews of Modern Physics* **78**, 17-85 (2006).
25 Keimer, B., Kivelson, S. A., Norman, M. R., Uchida, S. & Zaanen, J. From quantum matter to high-temperature superconductivity in copper oxides. *Nature* **518**, 179-186 (2015).
26 Tranquada, J. M. *et al.* Coexistence of, and Competition between, Superconductivity and Charge-Stripe Order in La1.6-xNd0.4SrxCuO4. *Physical Review Letters* **78**, 338-341 (1997).

27 Vinograd, I. *et al.* Competition between spin ordering and superconductivity near the pseudogap boundary in La2-xSrxCuO4: Insights from NMR. *Physical Review B* **106**, 054522 (2022).

28 Xu, G. *et al.* High-energy magnetic excitations from dynamic stripes in La1.875Ba0.125CuO4. *Physical Review B* **76**, 014508 (2007).

29 Parker, C. V. *et al.* Fluctuating stripes at the onset of the pseudogap in the high-Tc superconductor Bi2Sr2CaCu2O8+x. *Nature* **468**, 677-680 (2010).

30 Mak, K. F. & Shan, J. Simulating the Hubbard model with moiré semiconductors. *National Science Review* **13**, nwag069 (2026).

31 Li, H. *et al.* Imaging two-dimensional generalized Wigner crystals. *Nature* **597**, 650-654 (2021).

32 Jiang, Y. *et al.* Charge order and broken rotational symmetry in magic-angle twisted bilayer graphene. *Nature* **573**, 91-95 (2019).

33 Cao, Y. *et al.* Nematicity and competing orders in superconducting magic-angle graphene. *Science* **372**, 264-271 (2021).

34 Rubio-Verdú, C. *et al.* Moiré nematic phase in twisted double bilayer graphene. *Nature Physics* **18**, 196-202 (2022).

35 Wu, F., Lovorn, T., Tutuc, E. & Macdonald, A. H. Hubbard Model Physics in Transition Metal Dichalcogenide Moiré Bands. *Physical Review Letters* **121**, 026402 (2018).

36 Tang, Y. *et al.* Simulation of Hubbard model physics in WSe2/WS2 moiré superlattices. *Nature* **579**, 353-358 (2020).

37 Regan, E. C. *et al.* Mott and generalized Wigner crystal states in WSe2/WS2 moiré superlattices. *Nature* **579**, 359-363 (2020).

38 Xu, Y. *et al.* Correlated insulating states at fractional fillings of moiré superlattices. *Nature* **587**, 214-218 (2020).

39 Ciorciaro, L. *et al.* Kinetic magnetism in triangular moiré materials. *Nature* **623**, 509-513 (2023).

40 Tao, Z. *et al.* Observation of spin polarons in a frustrated moiré Hubbard system. *Nature Physics* **20**, 783-787 (2024).

41 Zhang, F. *et al.* Experimental signature of layer skyrmions and implications for band topology in twisted WSe2 bilayers. *Nature Physics* **21**, 1217-1223 (2025).

42 Pan, H., Wu, F. & Das Sarma, S. Band topology, Hubbard model, Heisenberg model, and Dzyaloshinskii-Moriya interaction in twisted bilayer WSe2. *Physical Review Research* **2**, 033087 (2020).

43 Zang, J., Wang, J., Cano, J. & Millis, A. J. Hartree-Fock study of the moiré Hubbard model for twisted bilayer transition metal dichalcogenides. *Physical Review B* **104**, 075150 (2021).

44 Bi, Z. & Fu, L. Excitonic density wave and spin-valley superfluid in bilayer transition metal dichalcogenide. *Nature Communications* **12**, 642 (2021).

45 Chubukov, A. V. & Varma, C. M. Quantum criticality and superconductivity in twisted transition metal dichalcogenides. *Physical Review B* **111**, 014507 (2025).

46 Wang, L. *et al.* Correlated electronic phases in twisted bilayer transition metal dichalcogenides. *Nature Materials* **19**, 861-866 (2020).

47 Foutty, B. A. *et al.* Mapping twist-tuned multiband topology in bilayer WSe2. *Science* **384**, 343-347 (2024).

48 Devakul, T., Crépel, V., Zhang, Y. & Fu, L. Magic in twisted transition metal dichalcogenide bilayers. *Nature Communications* **12**, 6730 (2021).
49 Wu, F., Lovorn, T., Tutuc, E., Martin, I. & Macdonald, A. H. Topological Insulators in Twisted Transition Metal Dichalcogenide Homobilayers. *Physical Review Letters* **122**, 086402 (2019).
50 Fernandes, R. M. & Venderbos, J. W. F. Nematicity with a twist: Rotational symmetry breaking in a moiré superlattice. *Science Advances* **6**, eaba8834 (2020).
51 Taillefer, L. Fermi surface reconstruction in high-Tc superconductors. *Journal of Physics: Condensed Matter* **21**, 164212 (2009).
52 Grüner, G. *Density Waves In Solids*. 1st edn, (CRC Press, 1994).
53 Grüner, G. The dynamics of charge-density waves. *Reviews of Modern Physics* **60**, 1129-1181 (1988).
54 Liu, G., Rumyantsev, S., Bloodgood, M. A., Salguero, T. T. & Balandin, A. A. Low-Frequency Current Fluctuations and Sliding of the Charge Density Waves in Two-Dimensional Materials. *Nano Letters* **18**, 3630-3636 (2018).
55 Walmsley, P. & Fisher, I. R. Determination of the resistivity anisotropy of orthorhombic materials via transverse resistivity measurements. *Review of Scientific Instruments* **88**, 043901 (2017).
56 Sengupta, S. *et al.* Transverse resistance due to electronic inhomogeneities in superconductors. *Physical Review B* **112**, 014514 (2025).

**Acknowledgement**
We thank Eugene Demler, Trithep Devakul, Rafael Fernandes, Aharon Kapitulnik, Eun-Ah Kim, Steven Kivelson, Patrick Lee, Wei-Cheng Lee, Allan MacDonald, Qimiao Si and Chandra Varma for helpful discussions.

# Figures

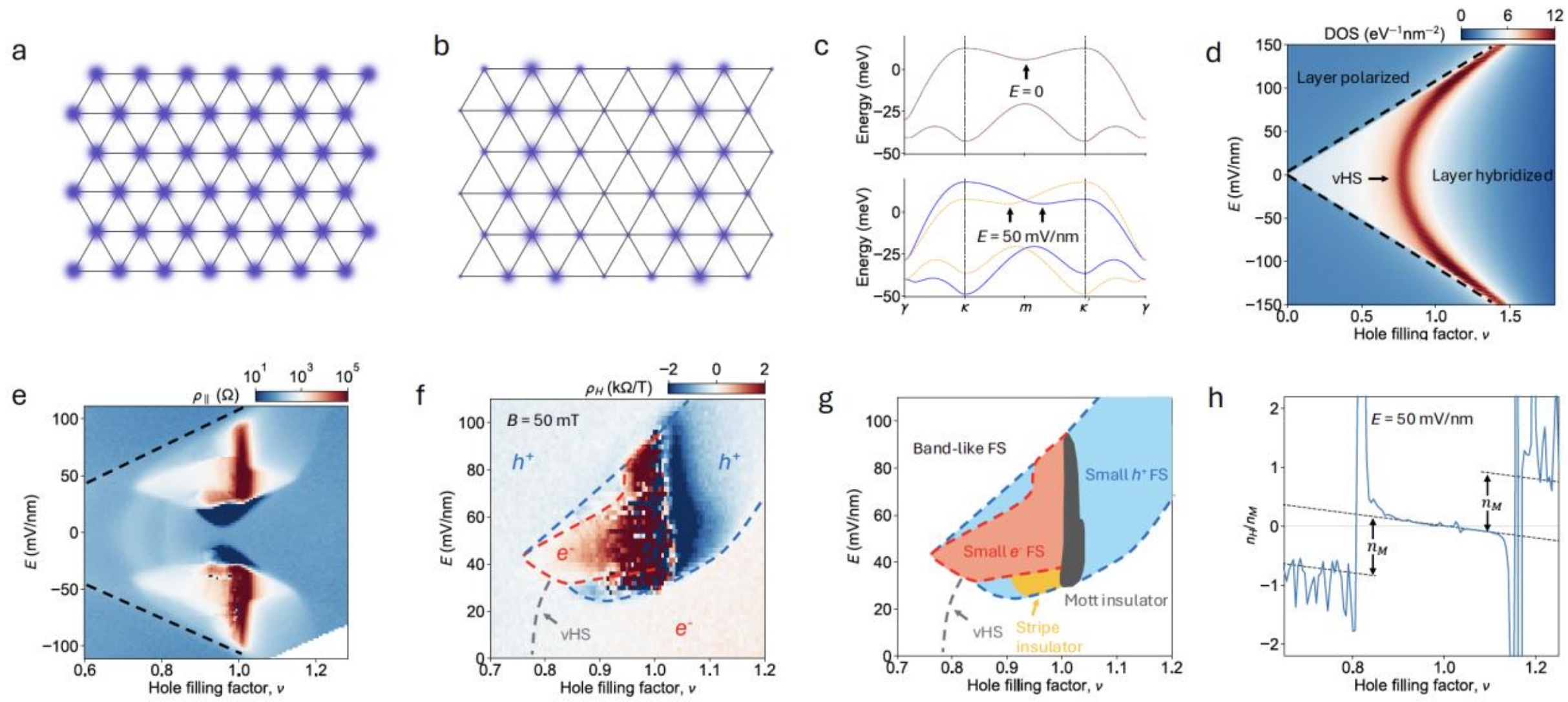

**Figure 1 | Fermi surfaces of a doped moiré Mott insulator. a,** Schematic moiré Mott insulator with one hole per moiré unit cell (i.e. $\nu = 1$) in tWSe2. **b,** Schematic stripe phase in a doped moiré Mott insulator. The hole density is self-organized into stripes with broken rotational symmetry and translational symmetry. **c,** The topmost moiré valence bands from the $K$-valley (spin up, blue) and $K'$-valley (spin down, orange) of tWSe2 at $E = 0$ (upper) and $E = 50\mathrm{mV/nm}$ (lower). Arrows mark the van Hove singularity (vHS). **d,** Electronic density of states (DOS) as a function of $\nu$ and $E$ showing the dispersion of the vHS. Results in **c** and **d** were obtained from continuum model calculations for a $4.0^0$ sample. **e,** Longitudinal resistivity $\rho_{\parallel}$ as a function of $\nu$ and $E$ at $B = 0\mathrm{T}$ (device 1, $3.9^0$ tWSe2). The black dashed lines in **d** and **e** mark the boundaries separating the layer-hybridized and -polarized regions. **f,** Zoom-in Hall coefficient $\rho_H$ map versus $\nu$ and $E$ measured at $B = 50\mathrm{mT}$. The sign of $\rho_H$ determines the dominant carrier type: $e^-$ and $h^+$ represent electron- and hole-type Fermi surface, respectively. Red and blue dashed lines mark the phase boundaries of the reconstructed Fermi surface of a doped Mott insulator. The grey line traces the vHS. **g,** Fermi surface phase diagram constructed from **f**. The grey and yellow regions mark the Mott and stripe insulator, respectively, where $\rho_H$ is not accessible; the orange and blue regions mark, respectively, the small electron ($e^-$) and hole ($h^+$) Fermi surfaces (FS) of a doped Mott insulator; the background white region marks the band-like FS. **h,** Hall density $n_H$ (normalized by the moiré density $n_M$) as a function of $\nu$ at constant $E = 50\mathrm{mV/nm}$. The dashed lines (from top to bottom) describe $\frac{n_H}{n_M} = -\nu + 2$, $\frac{n_H}{n_M} = -\nu + 1$ and $\frac{n_H}{n_M} = -\nu + 0$. A reset in $n_H$ by about $n_M$ is observed as the system transits from a doped Mott insulator with small FS to a band-like FS ($\nu \lesssim 0.8$ and $\nu \gtrsim 1.2$). All data were acquired at $T = 20\mathrm{mK}$.

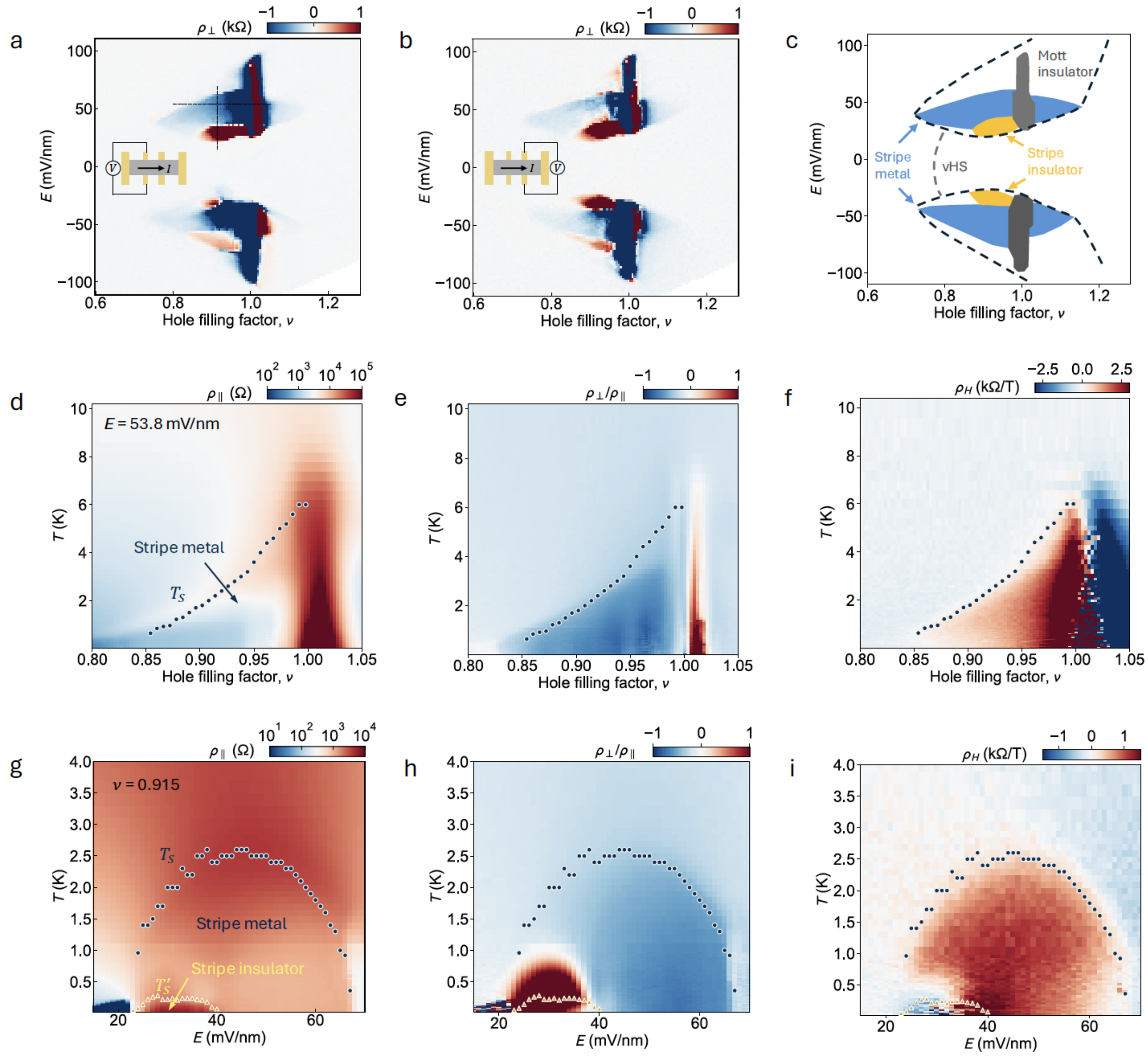

**Figure 2 | Emergent stripe phases from doping a Mott insulator. a, b,** Transverse resistivity $\rho_{\perp}$ as a function of $\nu$ and $E$ at $T = 20\text{mK}$ and $B = 0\text{T}$ measured for two different transverse electrode pairs. The inset shows the Hall bar measurement geometry. **c,** Constructed stripe phase diagram versus $\nu$ and $E$. The grey, blue and yellow regions mark the Mott insulator, the stripe metal and the stripe insulator, respectively. The dashed lines mark the vHS and the doped Mott insulator regions. **d-f,** $\rho_{\parallel}$ (**d**), $\rho_{\perp}/\rho_{\parallel}$ (**e**) and $\rho_H$ (**f**, measured at $B = 100\text{mT}$) as a function of $\nu$ and $T$ at constant $E = 53.8\text{mV/nm}$ (horizontal dashed line in **a**). The doping dependence of the stripe metal transition temperature $T_S$ for $\nu < 1$ is shown. The arrow denotes the stripe metal region. **g-i,** The same as **d-f** but as a function of $E$ and $T$ at constant $\nu = 0.915$ (vertical dashed line in **a**). The electric field dependence of $T_S$ and $T_S'$ (the stripe insulator transition temperature) are shown. The arrow denotes the stripe insulator region.

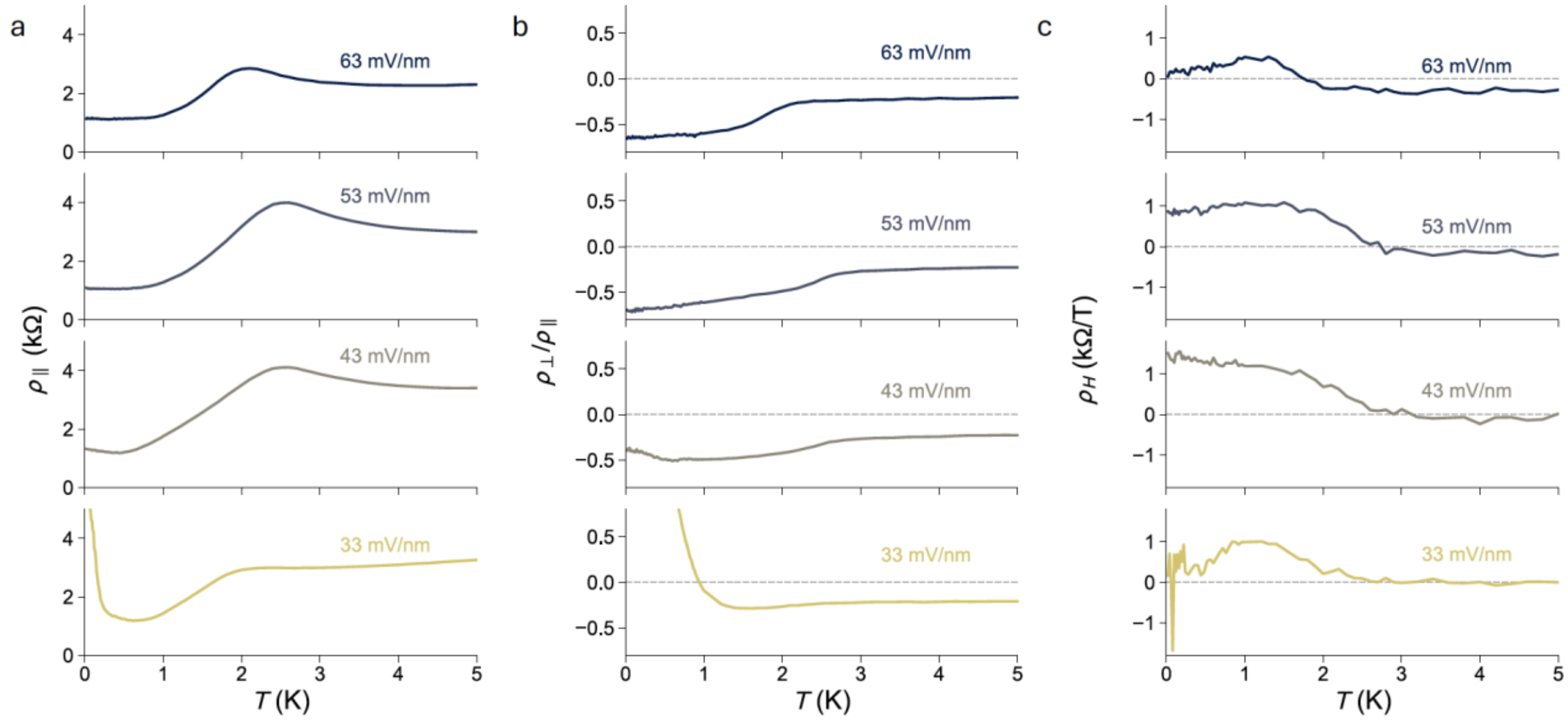

**Figure 3 | Thermodynamic phase transitions for stripe phases. a-c,** Temperature dependence of $\rho_{\parallel}$ (**a**), $\rho_{\perp}/\rho_{\parallel}$ (**b**) and $\rho_H$ (**c**, measured at $B = 100\text{mT}$) at $\nu = 0.915$ and selected electric fields $E = 33, 43, 53, 63\text{mV/nm}$. A peak/bump in $\rho_{\parallel}$ is observed at the stripe metal transition temperature $T_S$, and is correlated with the onset for $\rho_{\perp}/\rho_{\parallel}$ and $\rho_H$. At $E = 33\text{mV/nm}$, a second thermodynamic phase transition to a stripe insulator is observed at $T_S' \approx 0.24\text{K}$.

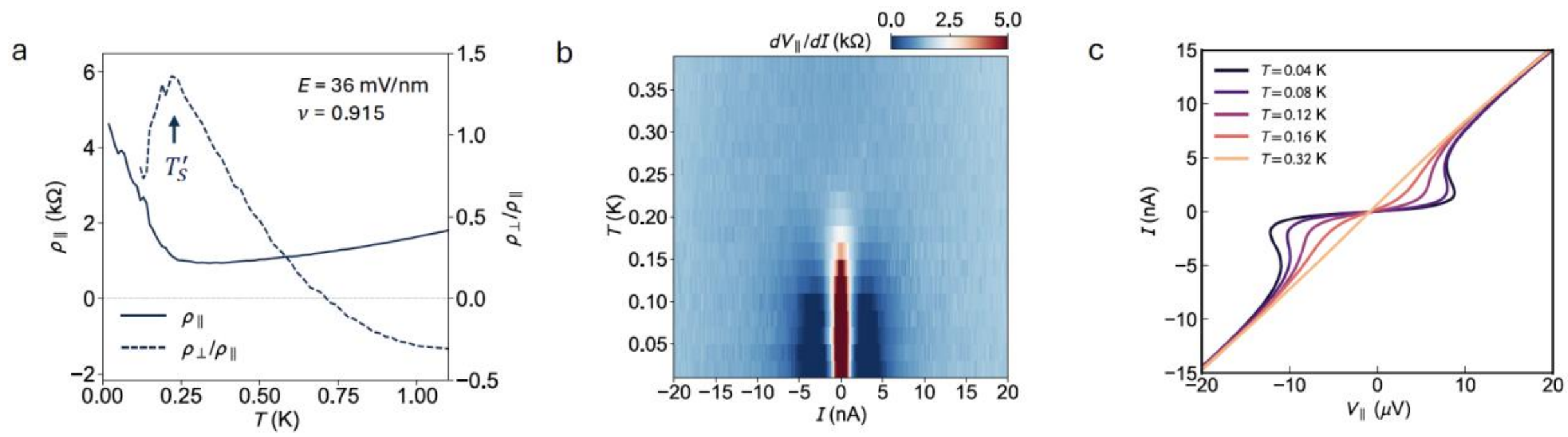

**Figure 4 | Insulating stripe phase and sliding density wave transport. a,** Temperature dependence of $\rho_\parallel$ (solid) and $\rho_\perp/\rho_\parallel$ (dashed) for an insulating stripe phase at $\nu = 0.915$ and $E = 36\mathrm{V/nm}$ ($B = 0$T). The arrow marks the stripe insulator transition temperature $T_S'$. **b,** Differential longitudinal resistance $\frac{dV_\parallel}{dI}$ as a function of bias current and temperature at the same $\nu$ and $E$. **c,** $I - V_\parallel$ curves at selected temperatures obtained by integrating $\frac{dV_\parallel}{dI}$ in **b** with respect to $I$.

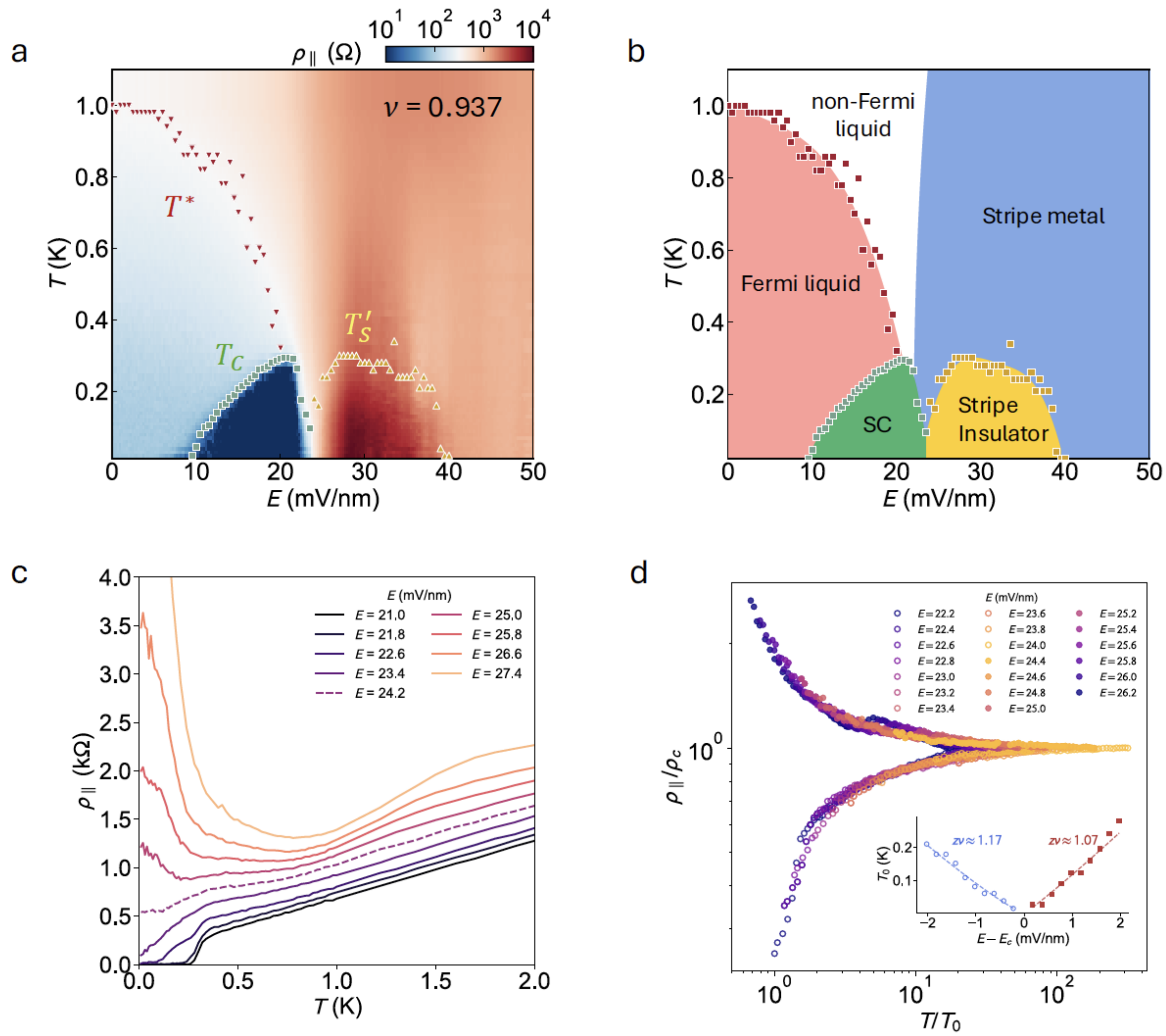

**Figure 5 | Stripe insulator-to-superconductor transition. a,** $\rho_{\parallel}$ as a function of $E$ and $T$ at $\nu = 0.937$ ($B = 0$T). The electric field dependence of the superconducting $T_C$ (green), $T_S'$ (yellow) and the Fermi liquid coherence temperature $T^*$ (red) are shown. **b,** Phase diagram constructed from **a** showing the Fermi liquid, non-Fermi liquid, superconductor (SC), stripe insulator and stripe metal. **c,** Temperature dependence of $\rho_{\parallel}$ at selected electric fields from **a**. The curve at $E_C \approx 24.2$mV/nm is shown as a dashed line. **d,** Scaling collapse of the resistivity curves near $E_C$ into two branches (see Methods for details). The inset shows the extracted $T_0$ as a function of $E - E_C$. The dashed lines are fits to the data by the power law $T_0 \propto |E - E_C|^{z\nu}$ on both the superconductor side (blue, $z\nu \approx 1.17$) and the stripe insulator side (red, $z\nu \approx 1.07$).

## Extended Data Figures

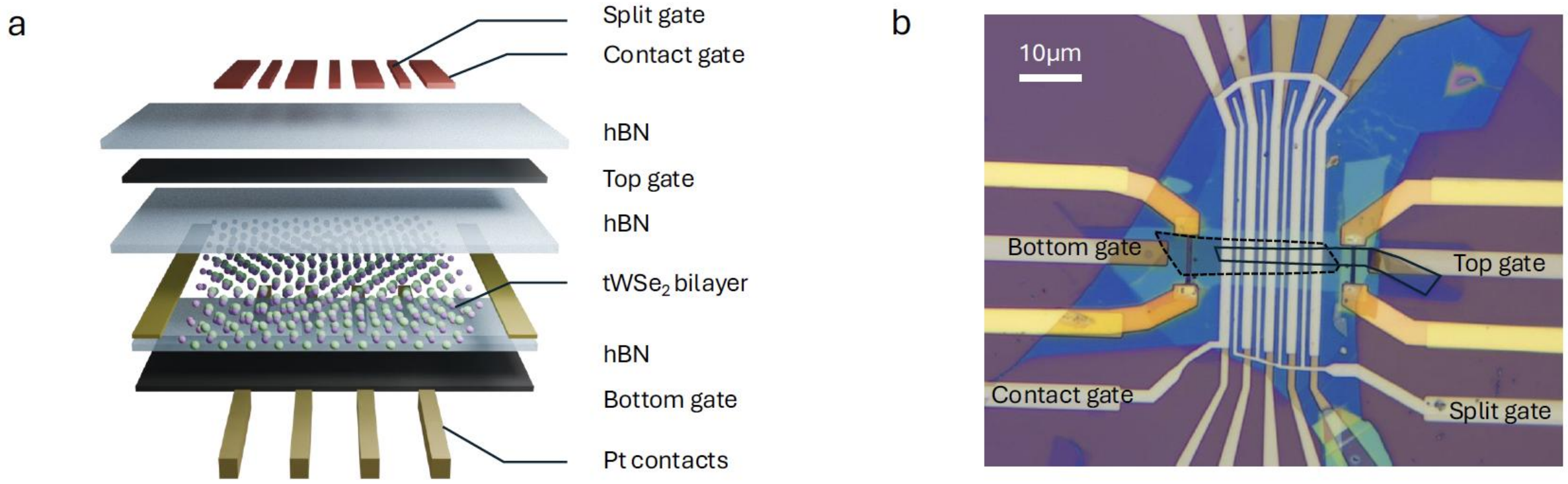

**Extended Data Figure 1 | Device structure. a,** Schematic device structure showing all the key elements. **b,** Optical micrograph of device 1 studied in the main text. The top and bottom graphite gates are denoted by solid and dashed lines, respectively. Their overlapped area defines the device channel. The scale bar is 10μm.

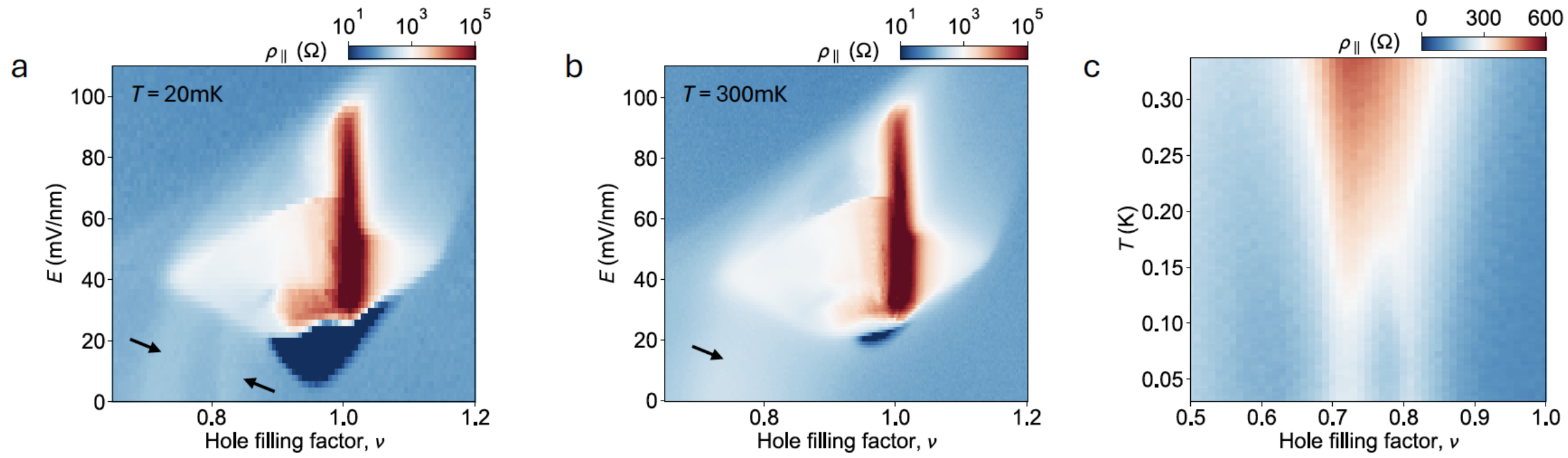

**Extended Data Figure 2 | Splitting of van Hove singularity at low temperatures. a, b,** $\rho_\parallel$ as function of $\nu$ and $E$ at $T = 20$mK (**a**) and $T = 300$mK (**b**) and at $B = 0$T. The arrows highlight the resistivity peaks near the vHS. **c,** $\rho_\parallel$ as function of $\nu$ and $T$ at $E = 0$mV/nm. Whereas a resistivity peak is observed at the vHS at high temperatures (also see **b**), the peak is split into two at temperatures below about 150mK, turning the resistivity to a dip at the vHS (also see **a**). The data may suggest a phase transition at around 150mK but the physics is beyond the scope of this study. Future works are required to understand the physical origin of the resistivity peak splitting at low temperatures.

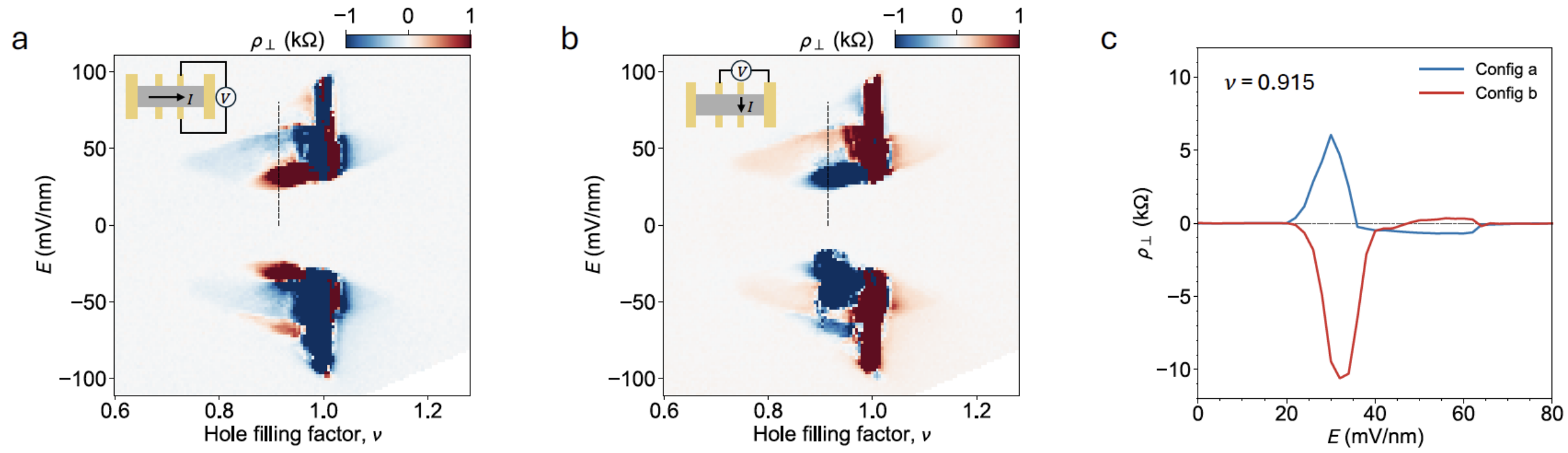

**Extended Data Figure 3 | Dependence of transverse resistivity on bias current direction. a, b,** $\rho_\perp$ as a function of $\nu$ and $E$ at $T = 20\text{mK}$ and $B = 0$ T for two different measurement configurations shown in the inset. The probe (or bias current) direction is rotated by 90 degrees. **c,** An illustration of two line cuts from **a** and **b** at constant $\nu = 0.915$ (along the vertical dashed lines in **a** and **b**), showing a sign change but of comparable magnitude under 90-degree rotation.

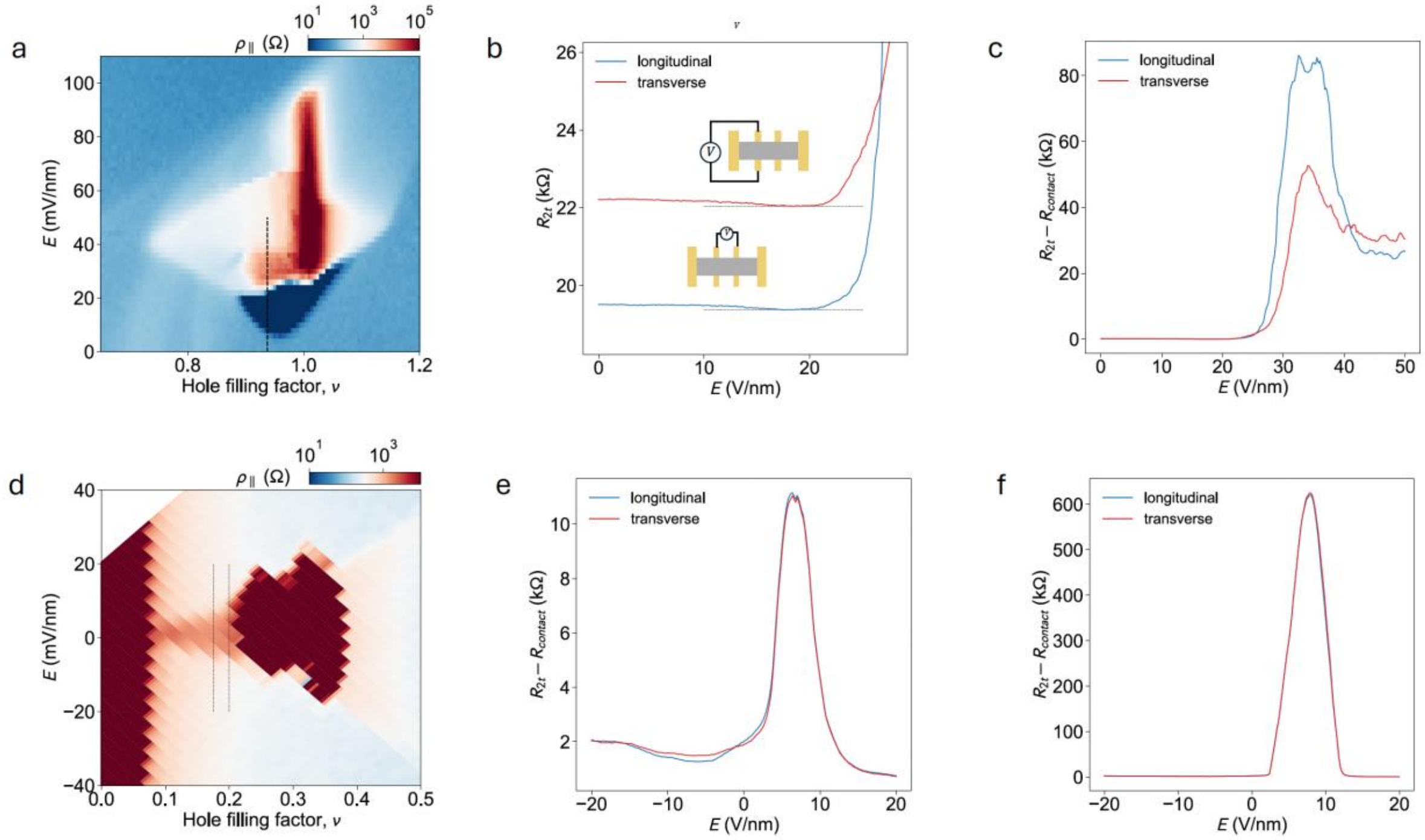

**Extended Data Figure 4 | Anisotropic two-terminal resistance. a, d,** $\rho_{\parallel}$ as a function of $\nu$ and $E$ at $T = 20$mK and $B = 0$T. Panel **a** focuses on the positive electric field side near $\nu = 1$. Panel **d** focuses on the low-filling region. **b,** Two-terminal resistance $R_{2t}$ measured along the longitudinal (blue) and transverse (red) direction (illustrated in insets) along the vertical dashed line in **a** ($\nu = 0.937$). The horizontal dashed lines mark the contact resistances ($R_{contact}$) at the superconducting state. **c,** Electric field dependence of the sample resistance $R_{2t} - R_{contact}$ (assuming a gate-independent contact resistance) corresponding to the two measurement configurations shown in **b**. Substantial transport anisotropy is observed in the doped Mott insulator region ($E \gtrsim 25$mV/nm). **e, f,** Electric field dependence of $R_{2t} - R_{contact}$ at $\nu = 0.175$ (**e**) and $\nu = 0.2$ (**f**) corresponding to the two measurement configurations shown in **b** (along the vertical dashed lines in **d**). In contrast to **c**, no obvious transport anisotropy is observed here although the sample resistance is strongly electric field dependent at both $\nu = 0.175$ and $\nu = 0.2$. The result rules out the electronic inhomogeneities scenario (see Methods) as local resistance variations would create transport anisotropy for all line cuts. Note that we did not scale $R_{2t} - R_{contact}$ for the two measurement configurations by any factor, suggesting the effective sample lengths along the longitudinal and transverse directions are nearly identical in the two-terminal measurements.

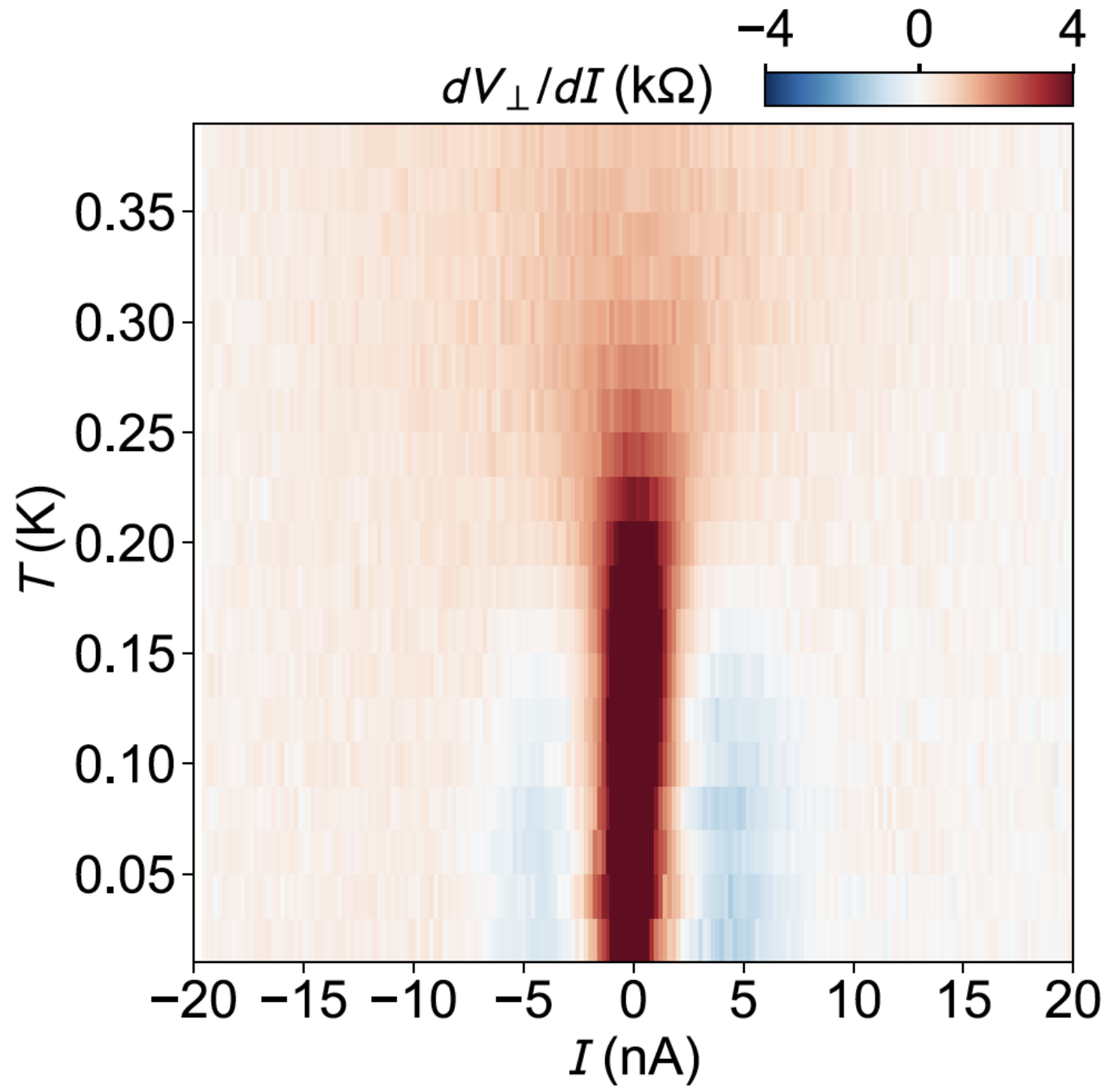

**Extended Data Figure 5 | Differential transverse resistance $dV_\perp/dI$ as a function of bias current and temperature at $E = 36$ mV/nm and $\nu = 0.915$.** Large zero-bias transverse resistance is observed for the stripe insulator below $T_S' \approx 200$mK. For $T < T_S'$, a small bias current exceeding 2nA quickly suppresses the transverse signal and depins the stripe insulator. The result perfectly correlates with the differential longitudinal resistance data shown in Fig. 4b. The positive $\frac{dV_\perp}{dI}$ at high bias currents show that the system remains in the stripe metal phase at high biases.

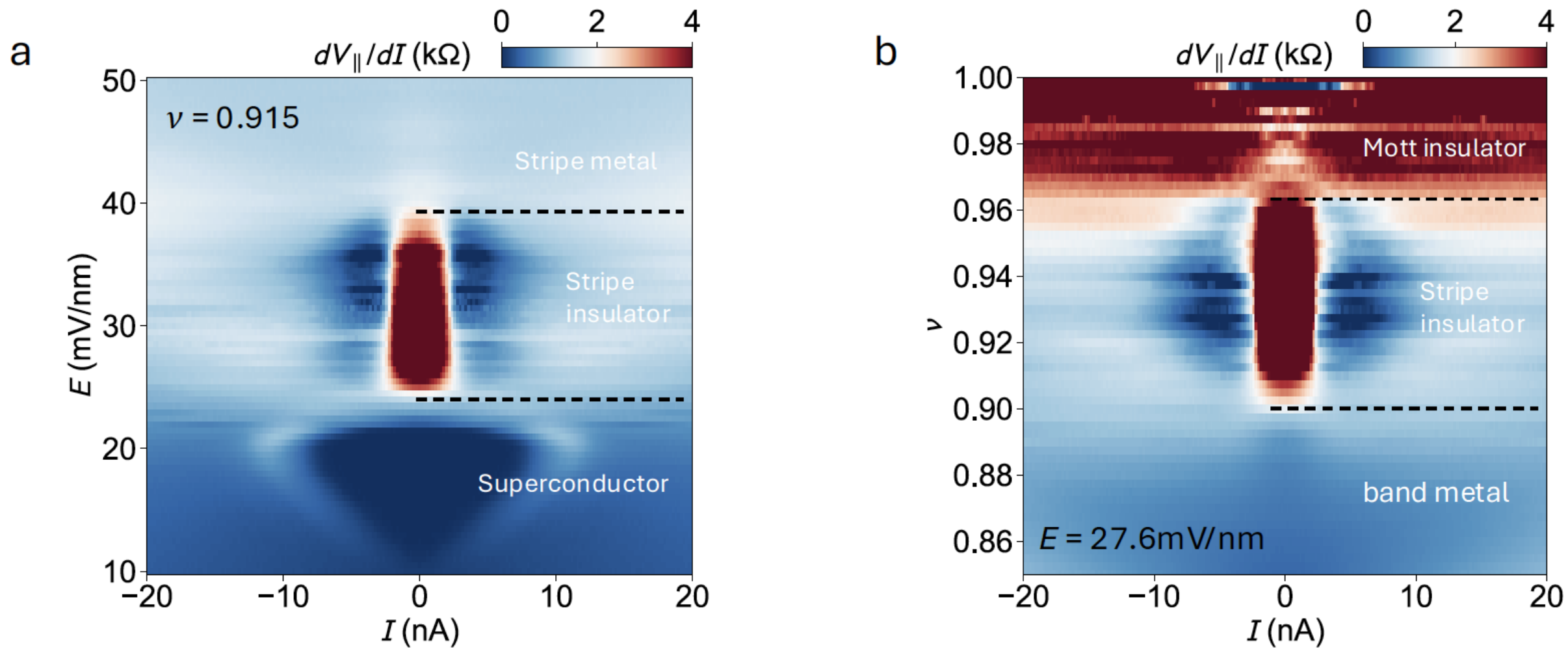

**Extended Data Figure 6 | Differential resistance $\frac{dV_\parallel}{dI}$ maps. a,** $\frac{dV_\parallel}{dI}$ as a function of $I$ and $E$ at constant $\nu = 0.915$. The electric field range covers the stripe metal, stripe insulator and superconductor, whose phase boundaries are marked by dashed lines. **b,** $\frac{dV_\parallel}{dI}$ as a function of $I$ and $\nu$ at constant $E = 27.6\text{mV/nm}$. The dashed lines separate the Mott insulator, stripe insulator and band metal. In both maps, both non-ohmic transport and negative differential resistance are observed for the stripe insulator, demonstrating sliding density wave transport.

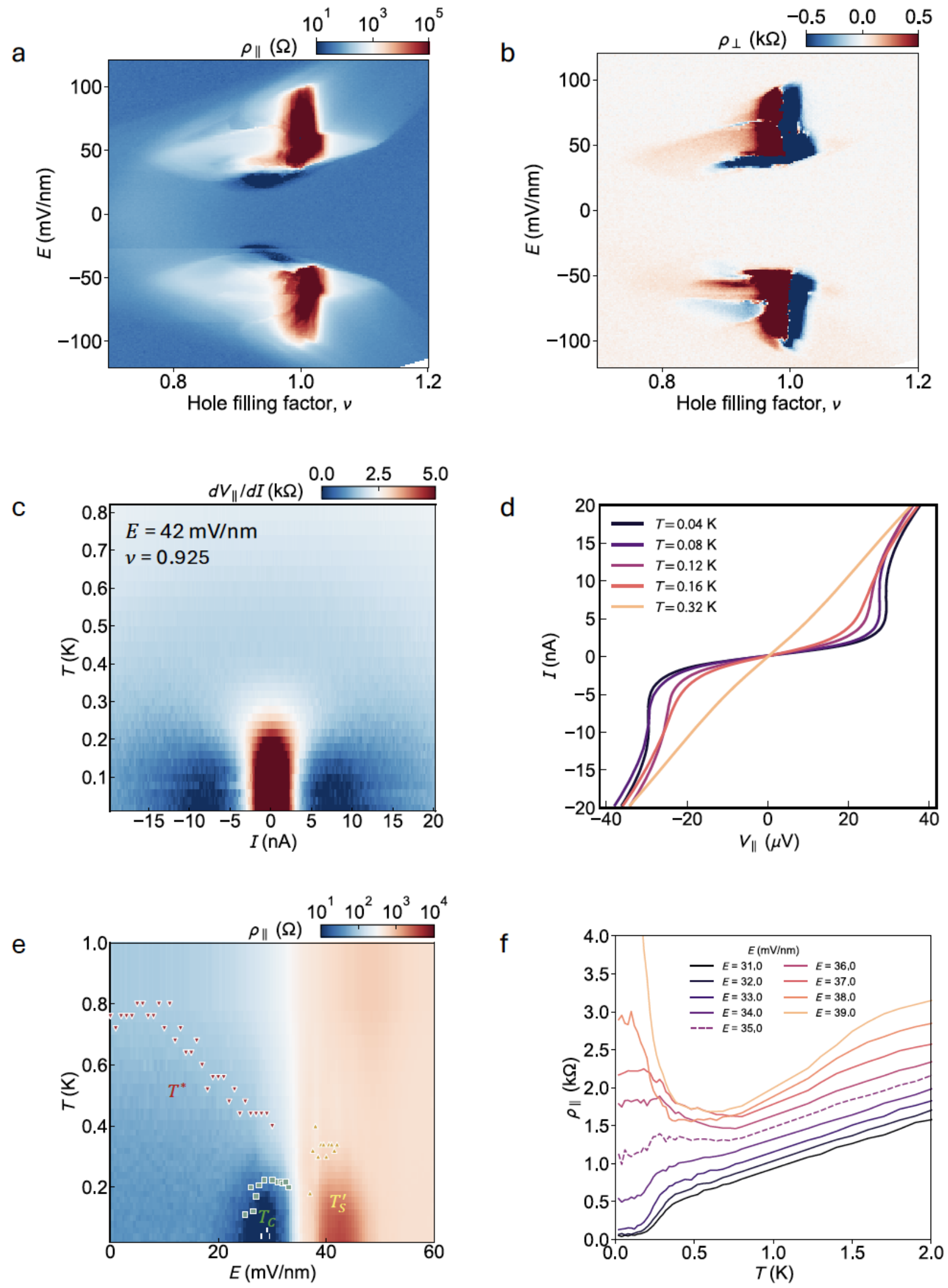

**Extended Data Figure 7 | Reproducibility in a $4.0^0$ device (device 2). a, b,** $\rho_\parallel$ (**a**) and $\rho_\perp$ (**b**) as a function of $\nu$ and $E$ at $T = 20\text{mK}$ and $B = 0\text{T}$. The phase diagram is very similar to that of device 1. Anisotropic transport is observed in the doped Mott insulator region, consistent with device 1. However, a different sign of $\rho_\perp$ is observed in the $E > 0$ region, showing that the stripe orientation (with respect to the Hall bar direction) in device 2 is different from that in device 1. **c,** $\frac{dV_\parallel}{dI}$ as a function of $I$ and $T$ for a stripe insulator at $E = 42\text{mV/nm}$ and $\nu = 0.925$. Both non-ohmic transport and negative differential resistance are observed, demonstrating sliding density wave transport. **d,** $I - V_\parallel$ curves at selected temperatures obtained by integrating $\frac{dV_\parallel}{dI}$ in **c** with respect to $I$. **e,** $\rho_\parallel$ as a function of $E$ and $T$ at $\nu = 0.925$. The electric field dependence of $T_C$ (green), $T_S'$ (yellow) and $T^*$ (red) are shown. **f,** Temperature dependence of $\rho_\parallel$ at selected electric fields from **e**. The curve at $E_C \approx 35.0\text{mV/nm}$ is shown as a dashed line.

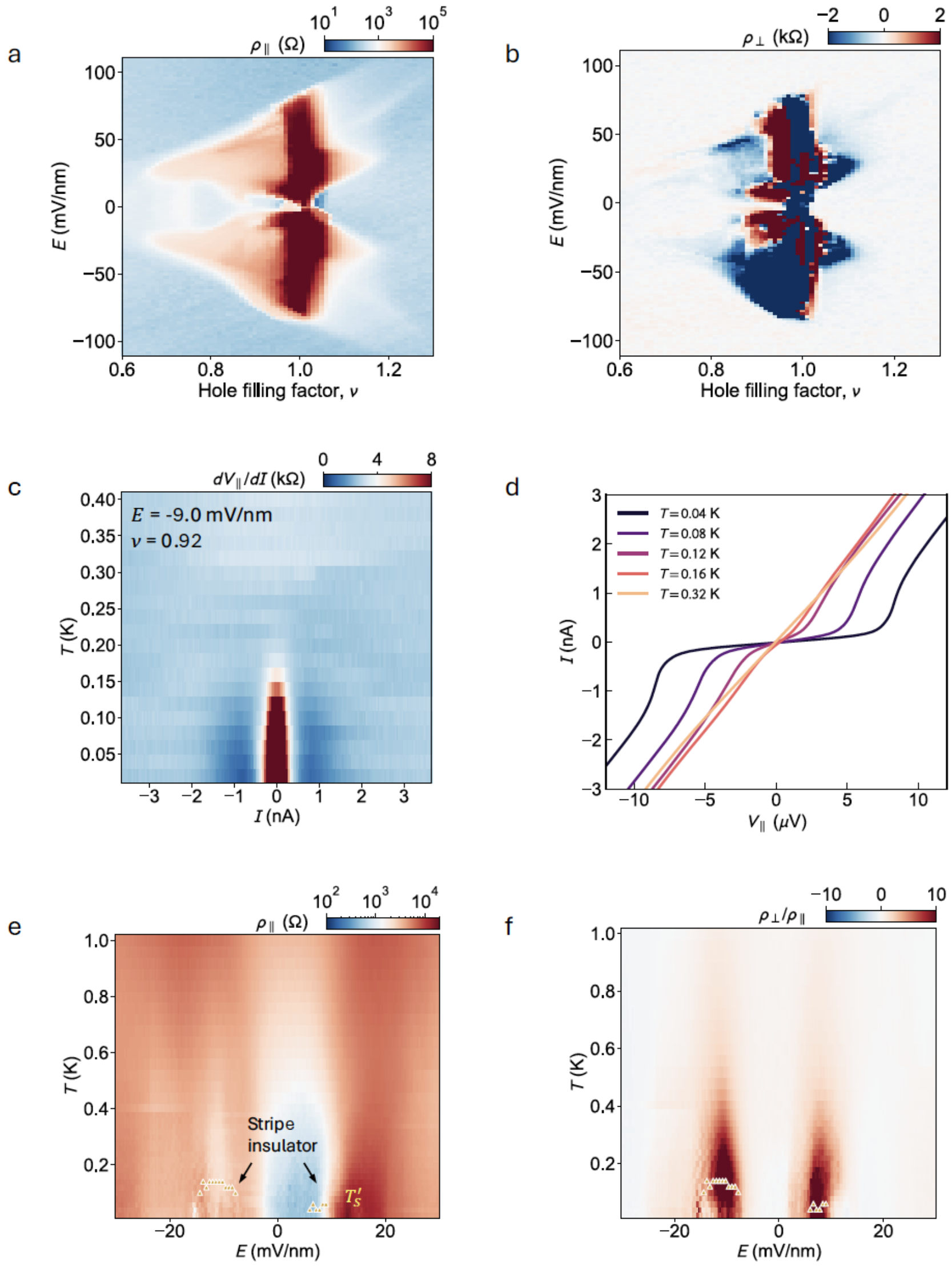

**Extended Data Figure 8 | Reproducibility in a $3.4^0$ device (device 3). a,** $\rho_{\parallel}$ as a function of $\nu$ and $E$ at $T = 20\text{mK}$ and $B = 0\text{T}$. Compared to the $4.0^0$ device, the electronic correlation effect is stronger, as manifested by a more robust Mott insulator and a larger area of the doped Mott insulator region surrounding $\nu = 1$. **b,** The corresponding $\rho_{\perp}$ as a function of $\nu$ and $E$. Again, anisotropic transport is observed in the doped Mott insulator region. **c,** $\frac{dV_{\parallel}}{dI}$ as a function of $I$ and $T$ for a stripe insulator at $E = -9.0\text{mV/nm}$ and $\nu = 0.92$. Non-ohmic transport characteristic of sliding density wave transport is observed. **d,** $I - V_{\parallel}$ curves at selected temperatures obtained by integrating $\frac{dV_{\parallel}}{dI}$ in **c** with respect to $I$. **e, f,** $\rho_{\parallel}$ (**e**) and $\rho_{\perp}/\rho_{\parallel}$ (**f**) as a function of $E$ and $T$ at constant $\nu = 0.92$. Superconductivity is absent in this doping density. The electric field dependence of $T_S'$ is shown. The arrows denote the stripe insulator region.

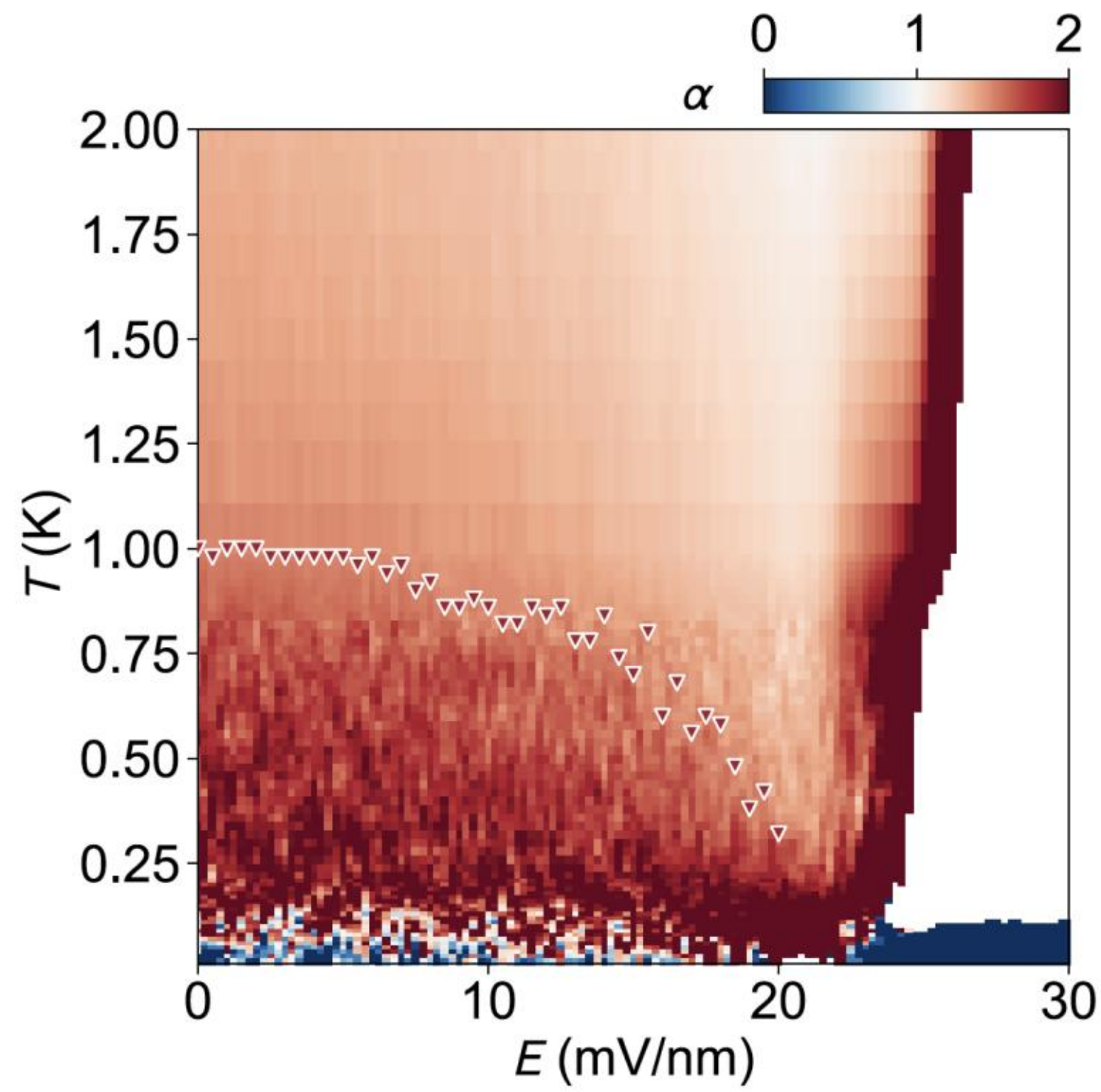

**Extended Data Figure 9 | Extracted temperature exponent $\alpha(T)$ as a function of $E$ and $T$ at $\nu = 0.937$.** A magnetic field $B = 100\text{mT}$ is applied to suppress superconductivity. The exponent $\alpha(T)$ is extracted from $\rho_{\parallel}(T) = \rho_0 + AT^{\alpha(T)}$ by fitting the local slope of $\ln(\rho_{\parallel} - \rho_0)$ versus $\ln T$ (Methods). $T^*$ (red triangles) is defined by the contour line $\alpha(T) = 1.5$, below which the system is a Fermi liquid, i.e. $\alpha \approx 2$.